\documentclass{emulateapj}
\usepackage{wasysym}
\usepackage{graphicx}
\usepackage{amssymb}
\usepackage{textcomp}

\shorttitle{Spin-orbit alignment in NY\,Cephei}
\shortauthors{Albrecht et al.}
\begin{document}

\title{
The Banana Project.~III.~Spin-orbit alignment in the long-period eclipsing binary NY\,Cephei\altaffilmark{*}
}

\author{
Simon Albrecht\altaffilmark{1},
Joshua N.~Winn\altaffilmark{1},
Joshua A.~Carter\altaffilmark{1},\\
Ignas A.~G.~Snellen\altaffilmark{2},
Ernst J.~W.~de Mooij\altaffilmark{2}
}

\affil{\altaffilmark{1}Department of Physics, and Kavli Institute for
  Astrophysics and Space Research,\\Massachusetts Institute of
  Technology, Cambridge, MA 02139, USA}

\affil{\altaffilmark{2}Leiden Observatory, Leiden University, Postbus
  9513, 2300 RA Leiden, The Netherlands}

\email{albrecht@space.mit.edu}

\altaffiltext {*}{ Based on observations made with {\it Sophie}, a
  high-resolution \'{e}chelle spectrograph on the 1.93-m telescope of
  the Observatoire de Haute-Provence.}

\begin{abstract}

  Binaries are not always neatly aligned. Previous observations of the
  DI\,Her system showed that the spin axes of both stars are highly
  inclined with respect to one another and the orbital axis. Here we
  report on a measurement of the spin-axis orientation of the primary
  star of the NY\,Cep system, which is similar to DI\,Her in many
  respects: it features two young early-type stars ($\sim6$Myr,
  B0.5V+B2V), in an eccentric and relatively long-period orbit
  ($e=0.48$, $P=15\fd3$). The sky projections of the rotation vector
  and the spin vector are well-aligned
  ($\beta_{\mathrm{p}}=2\pm4^{\circ}$), in strong contrast to
  DI\,Her. Although no convincing explanation has yet been given for
  the misalignment of DI\,Her, our results show that the phenomenon is
  not universal, and that a successful theory will need to account for
  the different outcome in the case of NY\,Cep.

\end{abstract}


\section{Introduction}
\label{sect:introduction}

We are conducting measurements of the relative orientations of the
rotational and orbital axes in close binary star systems. Our name for
this undertaking is the Banana project, an acronym chosen to remind us
that binaries are not always neatly aligned. This paper reports our
results for the NY\,Cephei system, the third system we have studied,
the first two having been V1143\,Cygni \citep{albrecht2007} and
DI\,Herculis \citep{albrecht2009}. Those two earlier works were
written before our project was enlarged and named, and are
retrospectively included in this series as Papers I and II. The goal
of the Banana project is to enlarge the number of detached binary
systems for which the relative orientation of the spin axes is known,
in order to shed light on the formation and evolution of binaries and
perhaps also of planets.

\subsection{Motivation}

Close binaries and star-planet systems might be expected to have
well-aligned orbital and spin angular momenta, since all of the
components trace back to the same portion of a molecular cloud.
However, good alignment is not guaranteed. If a cloud is highly
elongated, with its long axis tilted with respect to its rotation
axis, then the cloud may give birth to binary stars with a strong
spin-orbit misalignment \citep{bonnell1992}. Alternatively, disks
around young stars might become warped during the last stage of
accretion. This warp could torque the orbit by a large angle while
maintaining the orientation of the spins \citep{tremaine1991}. More
generally, star formation may be a chaotic process, with accretion
from different directions at different times \citep{bate2010}, and
perhaps we should not expect the angular momenta of the star and the
disk to be as well aligned as they apparently were in the Solar
system.

There are also processes that could alter the stellar and orbital spin
directions after their formation. A third body orbiting a close pair
on a highly inclined orbit can introduce large oscillations in the
orbital inclination and eccentricity of the close pair
\citep{kozai1962}. Tidal dissipation during the high-eccentricity
phases can cause the system to free itself of these ``Kozai
oscillations'' and become stuck in a high-obliquity state
\citep{fabrycky2007}. However, if dissipation is sufficiently strong
then the system will evolve into the double-synchronous state,
characterized by spin-orbit alignment \citep[e.g.][]{hut1981}.
Therefore, whether a close binary or a star-planet system is
well-aligned or misaligned depends on its particular history of
formation and evolution. Even though this issue is important for a
complete understanding of star formation, there has been very little
observational input.

\begin{table*}
  \begin{center}
    \caption{Measurements of spin orbit angles in eclipsing binaries\label{tab:angles}}
    \smallskip 
    \begin{tabular}{l  r@{$+$}l   r@{.}l  
        r@{$\pm$}l   c c  r}
      \tableline\tableline
      \noalign{\smallskip}
         System & \multicolumn{2}{c}{Spectral Type} & \multicolumn{2}{c}{Period [d]} 
         &  \multicolumn{2}{c}{$r_{\rm p}$}    & Eccentricity ($e$)  & $\beta$[$^{\circ}$] &     reference   \\
      \noalign{\smallskip}
      \hline
      \noalign{\smallskip}
      $\beta$\,Lyr    & Be&B6-8II   & 12&9   & \multicolumn{2}{c}{0.52}  & $<$0.01            &   aligned?                        & 1,2 \\
      V1010\,Oph    &    A7IV-V& ?    &  0&66   &   0.4511 &0.011                    & $\equiv$~0         &   aligned?                        &   3,4 \\
      W\,Umi          &   A3V&G9IV     &  1&7   &   0.363 &0.001                     & $\equiv$~0         &   aligned?                        &  5  \\       
      $\delta$\,Lib   & A0V&K0IV    & 2&33   &  0.30&0.06                      & $\equiv$~0         &   aligned?                        & 6,7  \\
      V505\,Sgr       & A2V&GIV     &  1&2   & 0.288&0.001                     & $\equiv$~0         &   aligned?                        & 8,9 \\
       AI\,Dra         & A0V&F9.5V   &  1&2   & 0.284&0.004                     & $\equiv$~0         &   aligned?                        & 8,10 \\
      X\,Tri       &   A3V& G3IV     &  0&97   &   0.272 &0.003                     & $\equiv$~0         &   aligned?                        &   5 \\
      RZ\,Cas     &  A3V &KIV     &  1&20   &   0.233 &0.001                     & $\equiv$~0         &   aligned?                        &  5 \\
      U\,Sge           &   B8.5V&G3III     &  3&38   &   0.219 &0.002                     & 0.04         &   aligned?                        &   5 \\
      WW\,Cyg           &   B8V&G4III     &  3&32   &   0.215 &0.002                     & $\equiv$~0         &   aligned?                 &  5 \\
      Y\,Leo          &   A3V&K3IV   &  1&69   &   0.213 &0.001                     & $\equiv$~0         &   aligned?                        &   5 \\
      RX\,Hya           &   A8&K0IV     & 2&28   &   0.211 &0.013                     & $\equiv$~0         &   aligned?                        &  5 \\    
      Algol           & B8&K2IV     & 2&87   & 0.206&0.003                     & $\equiv$~0         &   aligned?                        &   11,12,13 \\
      RW\,Gem           &   B5-B6V&F0III     &  2&87   &   0.198 &0.002                     & $\equiv$~0         &   aligned?                        &   5 \\
      Y\,Psc           &   A3V&K2IV     &  3&77   &   0.193 &0.001                     &  0.12        &   aligned?                        &   5 \\
      TV\,Cas          &   A2V&G1IV     &  1&81   &   0.188 &0.15                     & $\equiv$~0         &   aligned?                        &  5 \\
      ST\,Per           &   A3V&KIV     &  2&65   &   0.184 &0.004                     & $\equiv$~0         &   aligned?                        &  5 \\
      U\,Cep           &   B7V&G8III    &    2&49   & 0.177 &0.009                     & $\equiv$~0         &   aligned?                        &  5 \\
      TX\,Uma           &   B8V&F7-F8III    &  3&06   &   0.164 &0.002                     & $\equiv$~0         &   aligned?                 &  5 \\
      W\,Del          &   A0-B9.5V&K0IV     &  4&81   &   0.151 &0.004               & 0.20         &   aligned?                        &   5 \\
      AA\,Dor          &   sdOB& ?     &  0&26   &   0.14 &0.01               & $\equiv$~0         &   aligned?                        &   14 \\
      DE\,Dra         & B0V&B2V     &  5&3   & \multicolumn{2}{c}{0.14}  & 0.018$\pm$0.011    &   misaligned?                     & 15 \\
      SW\,Syg           &   A2V&K1IV     &  4&57   &   0.138 &0.002                     & 0.30        &   aligned?                        &   5 \\
      RY\,Per          &   B4V&F0III     &  6&86   &   0.137 &0.003                     & 0.21         &   aligned?                        &   5 \\
      RZ\,Sct          &   B2II&A0II-III    &  15&19  &  0.136 &0.007                     & $\equiv$~0         &   aligned?                        &   5 \\
      AQ\,Peg           &   A2V&K1IV     &  5&55   &   0.123 &0.002                     & 0.24         &   aligned?                        &   5 \\
      RY\,Gem          &   A2V&K0III-K1IV     &  9&30   &   0.097 &0.014               & 0.16         &   aligned?                        &   5 \\
      NY\,Cep         & B0V&B2V     & 15&3   & 0.086&0.015      & 0.445$\pm$0.004    &   $\beta_{\rm p} = 2^\circ \pm 4^\circ$ & 16,17 \\
      DI\,Her         & B4V&B5V     &10&55   &0.0621&0.001      & 0.489$\pm$0.003    &   $\beta_{\rm p} = 72^\circ \pm 4^\circ$,      $\beta_{\rm s} = -84^\circ \pm 8^\circ$   &  18,19 \\
      V1143\,Cyg      & F5V&F5V     & 7&64   & 0.059&0.001      & 0.5378$\pm$0.0003  &   $\beta_{\rm p} = 0\fdg 3 \pm 1\fdg 5$,      $\beta_{\rm s} = -1\fdg 2 \pm 1\fdg 6$  &  20,21 \\
      \noalign{\smallskip}
      \tableline
      \noalign{\smallskip}
      \noalign{\smallskip}
    \end{tabular}

    \tablecomments{In column 4, $r_{\rm p}$ denotes the radius of the
      primary star divided by the length of the orbital semimajor
      axis.  For DE\,Dra and $\beta$\,Lyr no uncertainties in the
      primary radii are given in the references.  In column 5, the
      entry ``$\equiv$~0'' indicates that the eccentricity was assumed
      to be zero by the authors. In column 6, $\beta_{\rm p}$and
      $\beta_{\rm s}$ denote angles between the projections of stellar
      and orbital spin axes for the primary and secondary,
      respectively, using the coordinate system of
      \citet{hosokawa1953}.}

    \tablerefs{ (1) \cite{rossiter1924};  (2) \cite{harmanec2002}; (3) 
      \cite{worek1988} (4) \cite{crcoran1991}; (5) \cite{twigg1979};
      (6) \cite{bakis2006}; (7) \cite{worek1985}; (8) \cite{worek1996}; (9)
      \cite{lzaro2006}; (10) \cite{lzaro2004}; (11) \cite{mclaughlin1924}; (12)
      \cite{struve1931}; (13) \cite{soderhjelm1980}; (14) \cite{rucinski2009}; (15)
      \cite{hube1982}; (16) \cite{holmgren1990}; (17) This study; (18)
      \cite{popper1982}; (19) \cite{albrecht2009}; (20)
      \cite{andersen1987}; (21) \cite{albrecht2007}
     }

  \end{center}
\end{table*}

Another motivation comes from exoplanetary science. Recently it was
revealed that many of the close-in giant planets (``hot Jupiters'')
have orbits that are strongly misaligned with the rotation axes of
their parent stars \citep[see, e.g.,][]{hebrard2008, winn2009,
  narita2009, triaud2010}. The high obliquities are especially common
among stars with higher masses and effective temperatures
\citep{winn2010,schlaufman2010}. The results of such studies are
commonly interpreted as constraints on theories of the ``migration''
processes that presumably brought the planets inward from their more
distant birthplaces. However, some theories invoke processes that
produce large stellar obliquities for reasons having nothing to do
with the planets, such as chaotic accretion \citep{bate2010}, and
magnetic interactions with the inner edge of the accretion disk
\citep{lai2010}. Although those theories have not been fully
developed, it would seem that the mechanisms they propose should also
operate in the case of binary stars, and therefore the theories might
be tested by measuring the obliquities of binary stars.

For example if interactions with a distant companion are important for
close double stars, then this should also be true for star-planet
systems. Other effects like tidal realignment do depend
  on the mass and mass ratio of the close pair. For a direct
  comparison with the exoplanet hosts, one would want to survey main
  sequence F and G binaries, while (as we will describe) most of the
  existing data, including the data presented in this paper, is for
  earlier type stars. We hope to rectify this situation with future
  observations.

\subsection{Measuring stellar obliquities}

Measuring the orientations of stellar spin axes is not
straightforward. Telescopes cannot usually resolve stellar disks,
causing the information on spatial orientation to be lost. An optical
interferometer, in combination with a high resolution spectrograph,
might allow for spatial resolution of rotationally-broadened stellar
absorption lines \citep{petrov1989, chelli1995}, but so far such
measurements are possible for only the very nearest and brightest
systems \citep{lebuquin2009}.

One would be able to determine the inclinations of the spin axes with
respect to the sky plane, based on empirical estimates of the
projected stellar rotation speed ($v\sin i$), the stellar radius
($R_\star$), and the stellar rotation period ($P_{\rm rot}$), using
the equation
\begin{equation}
i = \sin^{-1} \left[ \frac{v\sin i}{(2\pi R_\star/P_{\rm rot})} \right].
\end{equation}
Many investigators have pursued this path and found it to be blocked,
for reasons described clearly by \citet{soderblom1985}. Namely, the
quantities $v\sin i$, $R_\star$, and $P_{\rm rot}$ are difficult to
measure with high enough accuracy, and the flattening of the sine
function near $90^\circ$ prevents the method from discriminating even
modest inclinations from edge-on inclinations. Despite these
difficulties, \citet{glebocki1997} used this method to argue that
active binaries that are observed to be asynchronously rotating have
misaligned spin axes.

\begin{table}[t]
 \caption[Basic properties of NY\,Cep]{General data on NY\,Cephei}
 \label{tab:nycep}
 \begin{center}
 \smallskip
     \begin{tabular}{l l l}
	\hline
	\hline
	\noalign{\smallskip}
 	HIP & 113461  &  \\ 
       HD & 217312  &  \\ 
	R.A.$_{\rm J2000}$   & $22^{\rm h}58^{\rm m}40^{\rm s}$&\tablenotemark{$\ddagger$} \\
	Dec.$_{\rm J2000}$   & $63^{\circ}04^{\prime}38^{\prime\prime}$&\tablenotemark{$\ddagger$}  \\
	V$_{\mbox{max}}$     & $7.5$\,mag& \tablenotemark{$\ddagger$} \\
	Sp.\ Type           & B0.5V+ B2V &\tablenotemark{$\star$} \\
	Period             & $15\fd27$&\tablenotemark{$\dagger$}\\
        Eccentricity         & $0.48(2)$&\tablenotemark{$\star$} \\
	Inclination        & $78(1)^{\circ}$&\tablenotemark{$\star$} \\
        R$_{\rm p}$       & $6.8(7) R_{\odot}$&\tablenotemark{$\star$} \\
        R$_{\rm s}$       & $5.4(5) R_{\odot}$&\tablenotemark{$\star$} \\
        M$_{\rm p}$       & $13(1) M_{\odot}$&\tablenotemark{$\star$} \\
        M$_{\rm s}$       & $9(1) M_{\odot}$&\tablenotemark{$\star$} \\
	L$_{\rm s}$/L$_{\rm p}$     & 0.37(6) &\tablenotemark{$\star$}\\
	\noalign{\smallskip}
        \noalign{\smallskip}
	\hline
	\noalign{\smallskip}
        \noalign{\smallskip}
        $\ddagger$Data from \cite{esa1997}\\
        $\dagger$Data from \cite{ahn1992}\\
        $\star$Data from \cite{holmgren1990}
     \end{tabular}
     \tablecomments{ R$_{\rm p}$ denotes the radius of the primary
     component and R$_{\rm s}$ the radius of the secondary component.
     L$_{\rm p}$/L$_{\rm s}$ denotes the luminosity ratio between the
     primary and secondary.}
\end{center}
\end{table}

Given a large sample of $v\sin i$ measurements of stars that have a
known distribution of rotation speeds, or that are presumed to have
similar rotation speeds (e.g., main-sequence stars of a given mass and
age), it is possible to test for departures from an isotropic
distribution of spin directions.  Using this idea, \citet{guthrie1985}
and \citet{abt2001} searched for, and did not find, a tendency for
stars to be preferentially aligned with the Galactic plane. When the
orbital orientation is also known, as is the case for visual binaries
or eclipsing binaries, then such tests give information about
spin-orbit alignment.  Such studies are worthwhile but they are
hampered by the nonlinearity of the sine function (as mentioned above)
as well as uncertainty in the underlying distribution of rotational
speeds, and the heterogeneity in the techniques for measuring $v\sin
i$.  This method has been used in various guises by \citet{weis1974},
\citet{hale1994}, and \citet{howe2009}, among others.  Recently,
\citet{schlaufman2010} used this method on stars with transiting
planets, finding evidence that more massive stars have high
obliquities.

Indirect evidence on the orientation of young stars comes from the
orientation of their disks -- assuming star-disk alignment. The
projected rotation angle of the disk can be traced by the linear
polarization vector of the star light reflected from the dust grains
in disks \citep{monin1998}. These authors and others
\citep[e.g.][]{donar1999, wolf2001, jensen2004,monin2006} find that in
most binary systems the disks around the individual stars are aligned
with each other. However they also mention exceptions to this rule.
One of these exceptions is the hierarchical triple system T\,Tauri,
for which \cite{skemer2008} and \cite {ratzka2009} found that the disk
around the northern component, the original T\,Tauri system, is viewed
face on, while the disk around one of the southern components
(T\,Tau\,Sa), separated by $\gtrsim100$\,AU from T\,Tau\,N, is viewed
edge-on. Thus this system has misaligned circumstellar disks.

The method that provides the most accurate information for individual
systems takes advantage of eclipses. During an eclipse of one star by
another star or a planet, part of the rotating stellar surface is
hidden, causing a weakening of the corresponding velocity component of
the stellar absorption lines. Modeling of this spectral distortion
reveals the relative orientation of the spin and orbital axes on the
sky: the {\it projected obliquity}. This ``rotation anomaly'' was
first predicted by \cite{holt1893}.  A claim of its detection in the
$\delta$\,Librae system was made by \cite{schlesinger1910}, but more
definitive measurements were achieved by \citet{rossiter1924} and
\citet{mclaughlin1924} for the $\beta$\,Lyrae and Algol systems,
respectively. The phenomenon is now known as the Rossiter-McLaughlin
(RM) effect. Various aspects of the theory of the effect have been
worked out by \citet{struve1931, hosokawa1953, kopal1959, sato1974,
  otha2005, gimenez2006, hadrava2009}, and \citet{hirano2010}. The
work described in this paper, as well as in the previous papers in
this series, is based on the RM effect.

\subsection{Previous observations of the RM effect}

Although it has been more than 80 years since its discovery, there are
relatively few quantitative analyses of the RM effect observed in
stellar binaries. It is ironic that at present, there are more such
papers about exoplanetary systems than about stellar binaries. In the
past, observing the RM effect was generally either avoided (as a
hindrance to measuring accurate spectroscopic orbits) or used to
estimate stellar rotation speeds.  Almost all authors explicitly or
implicitly assumed that the orbital and stellar spins were aligned.

\begin{figure} [t]
  \begin{center}
   \includegraphics[width=8.9cm]{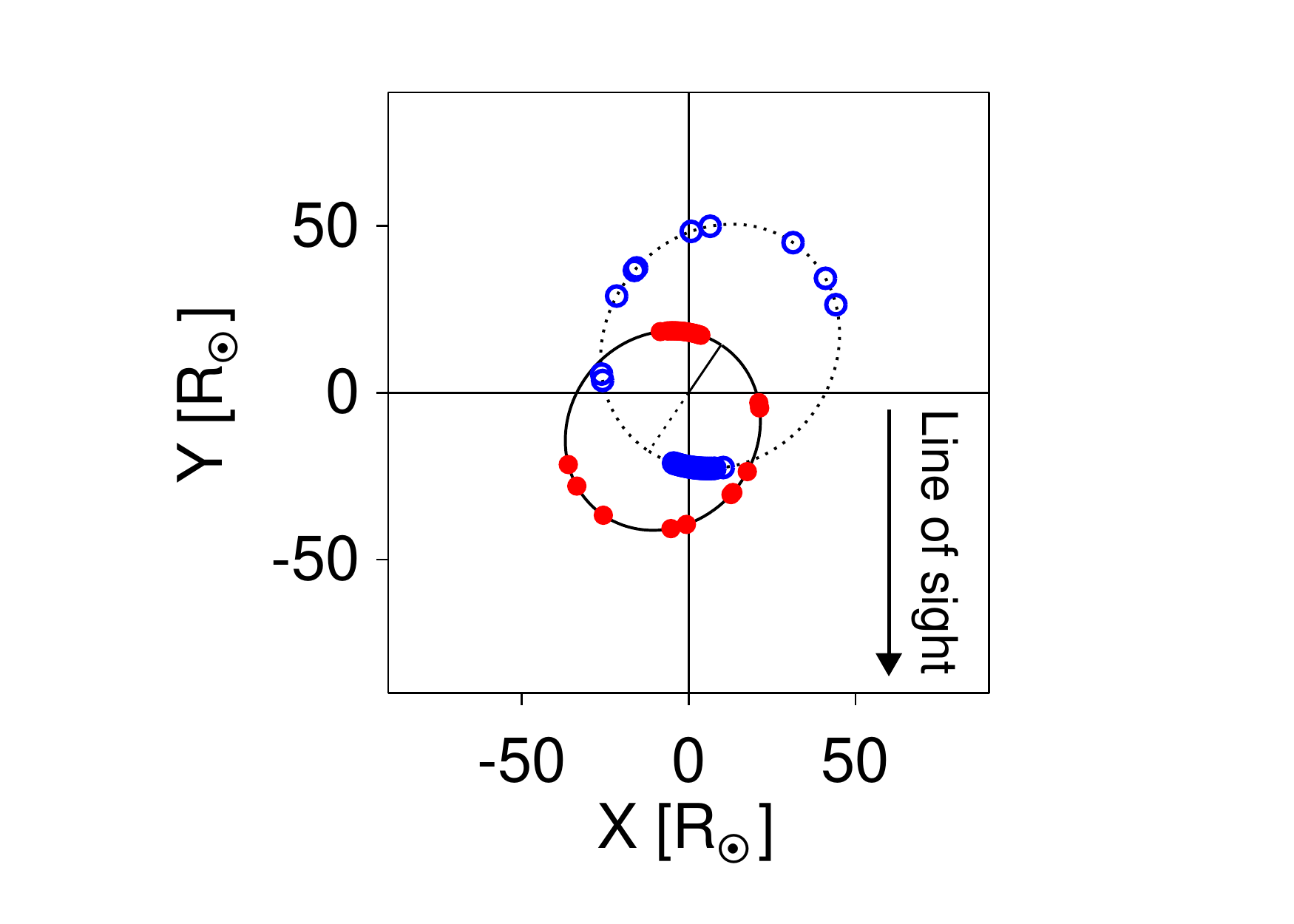}
   \caption {\label{fig:orbit_plane} {\bf Pole-on view of the orbit of
       NY\,Cep.} The solid and dashed ellipses indicate the orbits of
     the primary and secondary stars, respectively. The filled (red)
     circles indicate the positions of primary star and the open
     (blue) circles indicate the position of the secondary star
     during times of observations. The lines connecting the orbits
     with the center of gravity indicate the position of
     periastron. We observe the system from the $x$ origin and a
     negative $y$ value.}
  \end{center}
\end{figure}

Table~\ref{tab:angles} shows the results our literature search for
analyses of the RM effect in stellar binaries. (This table also gives
the results for NY\,Cep, the subject of this paper.)  We have tried to
be comprehensive, but cannot claim our list to be complete. The authors
would appreciate being notified of any omissions. The systems are
listed in order of decreasing $r_{\rm p}$, the radius of the
primary star in units of the orbital semimajor axis, and thus the
ordering is approximately a progression from closely-interacting
systems to well-detached systems (although this is not true for some
of the Algol systems, which have large faint secondaries). Column 6
summarizes the results for the stellar obliquities. In most cases we
have written ``aligned?'' because the RM data appear visually to
display the pattern of a well-aligned system---a redshift during the
first half of the eclipse, followed by a blueshift of equal amplitude
during the second half of the eclipse---with the question mark
indicating that no quantitative analysis was undertaken, and hence the
uncertainty is unknown. For systems with orbital inclinations very
close to 90$^\circ$ the results are especially ambiguous because in
such cases there is a strong degeneracy between the projected
obliquities and rotation rates \citep{gaudi2007}. For DE\,Dra we wrote
``misaligned?''  because \citet{hube1982} found the RM effect to be
asymmetric about the mideclipse time, but gave no quantitative result
for the projected obliquity. The only cases where quantitative results
for the projected obliquities are given are from the Banana project.

\begin{figure} [t]
  \begin{center}
   \includegraphics[width=8.5cm]{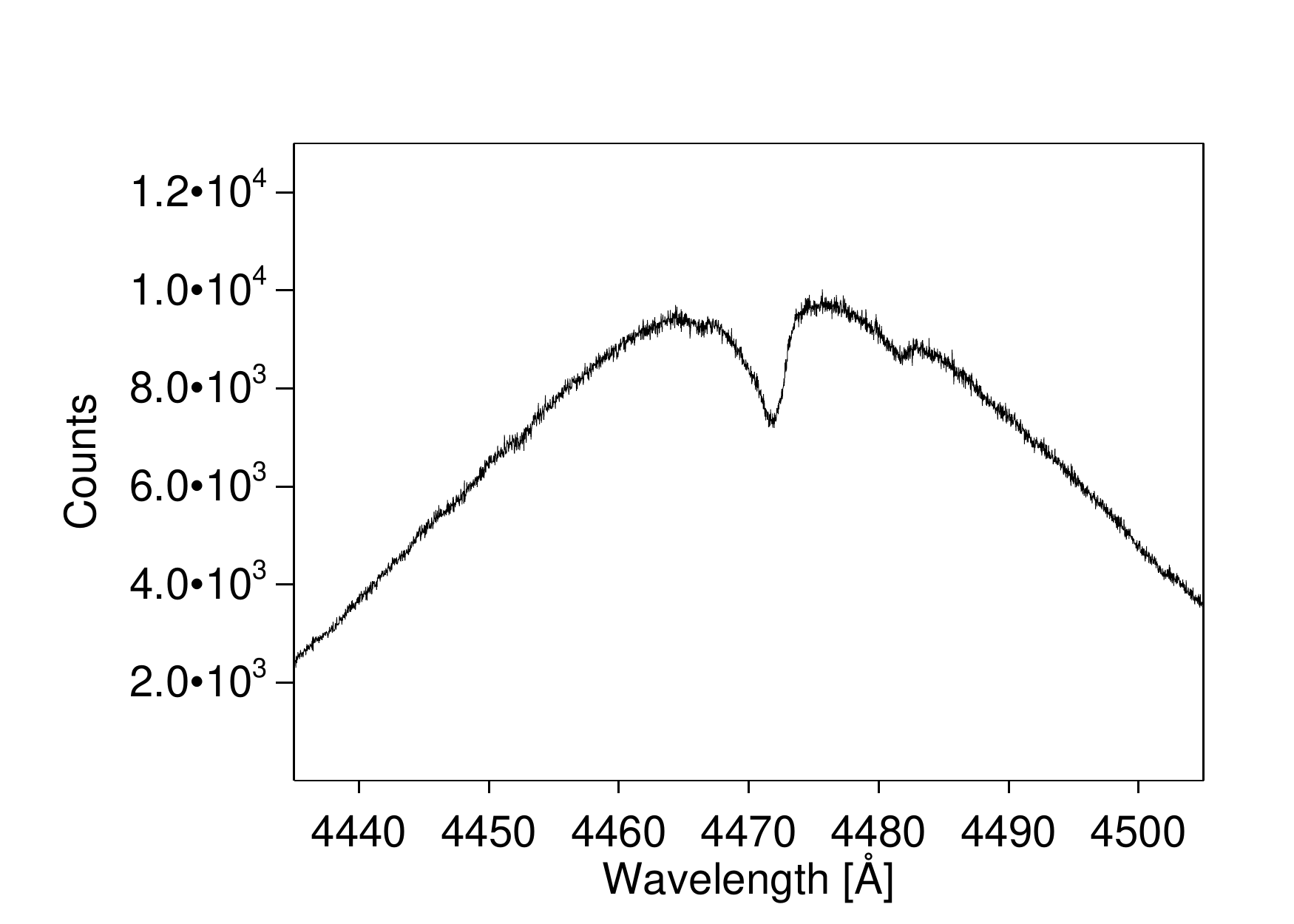}
   \caption {\label{fig:nycep_spec} {\bf Example spectrum of NY\,Cep.}
     Displayed is order number 11 of one spectrum of NY\,Cep, as
     delivered by the instrument software. No flat fielding, binning,
     or any other modifications to the data have been applied. The
     strong feature at 4471\,\AA{} is the He\,I line. Adjacent to this
     line, at 4481\,\AA{}, is the Mg\,II line.}
  \end{center}
\end{figure}

\subsection{The Banana project and NY\,Cep}

For our program, we have begun by concentrating on relatively young,
detached systems with $r_{\rm p} < 0.15$, in order to minimize the
effects of tidal interaction, and thereby study a more ``primordial''
distribution of obliquities.  We also omit Algol-type systems, to
avoid the extra complexity of the spin evolution caused by mass
transfer between the stars.  These criteria exclude the Algol-type
systems studied by \citet{twigg1979} as well as all the systems listed
in Table~\ref{tab:angles} except for DE\,Dra, V1143\,Cyg, DI\,Her, and
NY\,Cep.  The result of misalignment in the DE\,Dra system by
\citet{hube1982} is intriguing but needs to be checked.  For
V1143\,Cygni we showed that both spin axes are well-aligned with the
orbital axis [Paper I; \citet{albrecht2007}]. In contrast, for our
second target, DI\,Her, we measured spin axes that are drastically
misaligned with the orbital axis [Paper II;
\citet{albrecht2009}]. This was the first clear demonstration of such
a strong misalignment in a close binary. It resolved the longstanding
problem of the system's anomalous apsidal motion.

\begin{figure} [t]
  \begin{center}
   \includegraphics[width=8.9cm]{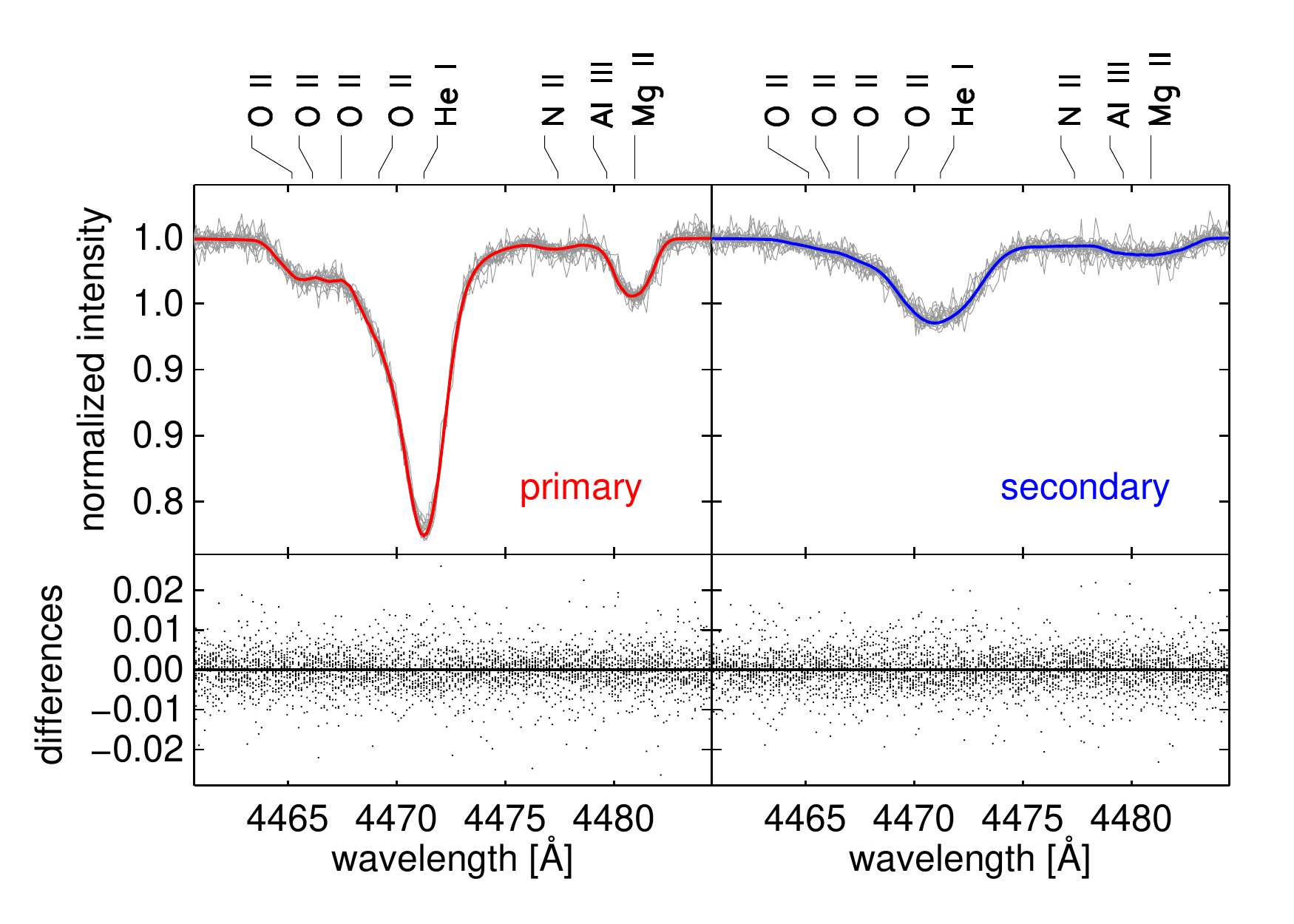}
   \caption {\label{fig:lines} {\bf Disentangled spectra of the two
       stars in the NY\,Cep system.} The two upper panels show spectra
     between 4461\,\AA{} and 4484\,\AA{}. The upper left panel is
     based on the sum of all the spectra obtained outside of the
     eclipse, after shifting into the rest frame of the primary star
     and after subtracting our best-fitting model of the secondary
     spectrum. The data are shown as gray lines, and the model as a
     continuous curve. The wavelengths of the lines used to construct
     the model spectrum are indicated above the panel. The lower left
     panel shows the difference between the data and the model.  The
     right panels show the secondary spectrum, after subtracting the
     best-fitting model of the primary spectrum.  For examples of the
     observed spectra with both stars present, see
     figures~\ref{fig:shape_out_of_eclipse} and
     \ref{fig:shape_primary_eclipse}.}
  \end{center}
\end{figure}

In this paper we focus on the NY\,Cephei system, whose properties are
summarized in Table~\ref{tab:nycep}. A summary of the history of
observations of this system was given by \citet{ahn1992}. It harbors
two early B type stars on an eccentric $15\fd27$ orbit. One peculiar
characteristic of this system is the lack of secondary
eclipses. Inferior conjunction occurs when the stars are relatively
far apart, and with an orbital inclination of $\apprle$$80^{\circ}$
the sky-projected separation of the stars is too large for
eclipses. The orientation of the orbit is illustrated in
Figure~\ref{fig:orbit_plane}.

Section~\ref{sect:observations} of this paper presents our
observations and  describes our model for the spectroscopic
data, and our analysis procedure.  The results for the orbital and
stellar parameters, including the measurement of the stellar
orientation are presented in section~\ref{sect:results}. The results
are discussed in section~\ref{sect:discussion}.

\section{Observations and Data Reduction}
\label{sect:observations}

We observed NY\,Cep with {\it Sophie}, a high-resolution \'{e}chelle
spectrograph on the 1.93-m telescope of the Observatoire de
Haute-Provence \citep{perruchot2008}, employing its High Efficiency
(R~$\approx$~40,000) mode. We chose an integration time of $20$\,min
resulting in signal-to-noise ratios (SNR) between $50$ and $150$ per
pixel at wavelengths near $4,500$\,\AA{}, the spectral region used for
our analysis. To illustrate the data quality, Figure
\ref{fig:nycep_spec} displays the region of interest of one of the
spectra.

The first half of the primary eclipse was observed during the night of
2009~Sep.~11/12.  On that night, clouds prevented us from obtaining
pre-ingress observations. The second half of the primary eclipse, as
well as a number of spectra directly after egress, were observed
during the night of 2009~Oct.\,12/13. Strong wind and poor seeing
caused lower SNRs than we had anticipated.  In addition, we gathered
some spectra on days spread throughout September to November 2009, in
order to establish the spectroscopic orbit.  All together, $27$
spectra were obtained during primary eclipse, $8$ spectra were
obtained directly after the end of egress, and another $11$
observations at various orbital phases outside of eclipses, for a
total of 46 spectra.  We also attempted photometry of the eclipse
throughout autumn and winter of 2009/2010, but all our attempts were
foiled by bad weather.

The two-dimensional reduced spectra were examined\footnote{The
  2D-spectra can be obtained from the following web address: {\tt
    http://atlas.obs-hp.fr/sophie/}.}, and bad pixels were flagged and
omitted from subsequent analysis. The spectra were shifted in
wavelength to account for the radial-velocity of the observatory
relative to NY\,Cep. Initial flat fielding was performed using the
nightly blaze functions.  We adopted the wavelength solution delivered
by the spectrograph software, which has an uncertainty of order
1~m\,s$^{-1}$ and is negligible for our purposes.

\begin{figure} [t]
  \begin{center}
   \includegraphics[width=8.cm]{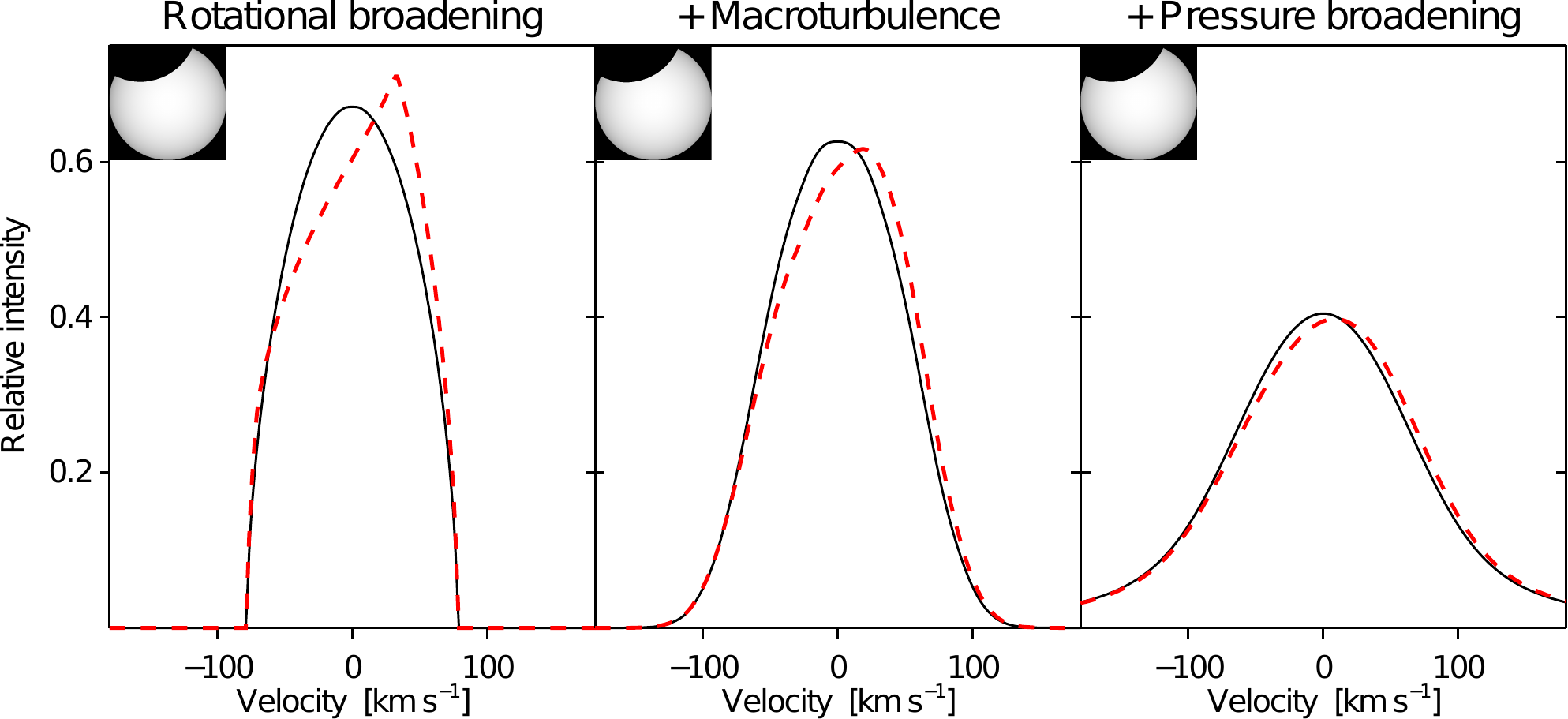}
   \caption {\label{fig:model_example} {\bf Illustration of the
       effects of rotation, macroturbulence, pressure broadening, and
       eclipse blockage on the absorption line model.} In each panel,
     the solid line represents an absorption line out of eclipse,
     while the (red) dashed line shows the same line just before mid-eclipse. The
     left panel shows only the effect of stellar rotation. The middle
     panel shows broadening by rotation and macro-turbulence (as
     appropriate for the Mg\,II line). The right panel shows a line
     additionally affected by pressure broadening (as appropriate for
     the He\,I line).}
  \end{center}
\end{figure}

\subsection{Description of the model} 
\label{sect:model}

Our analysis focused on the 11th \'{e}chelle order of the {\it Sophie}
CCD, which encompasses the wavelength range from 4459\,\AA{} to
4486\,\AA, including the two best absorption lines available for this
study, He\,I (4471\,\AA) and Mg\,II (4481\,\AA).  The helium line is
strong, with a line width dominated by pressure broadening. The
magnesium line is relatively weak but has the virtue of being chiefly
broadened by rotation and therefore well-suited to the analysis of the
RM effect. There are also a number of other weaker lines in this order
that must be modeled simultaneously with the stronger lines. The lines
are illustrated in Figure~\ref{fig:lines}, which shows the spectrum of
each star individually, after disentangling them with the modeling
procedure described in this section. The spectrum from that order was
binned to a resolution of about 9~km\,s$^{-1}$, giving 213 pixels in
the region of interest.  The SNR was estimated from the scatter in the
continuum on both sides of the He\,I (4471\,\AA) and Mg\,II
(4481\,\AA) lines.

\begin{figure}[t]
  \begin{center}
   \includegraphics[width=6.2cm]{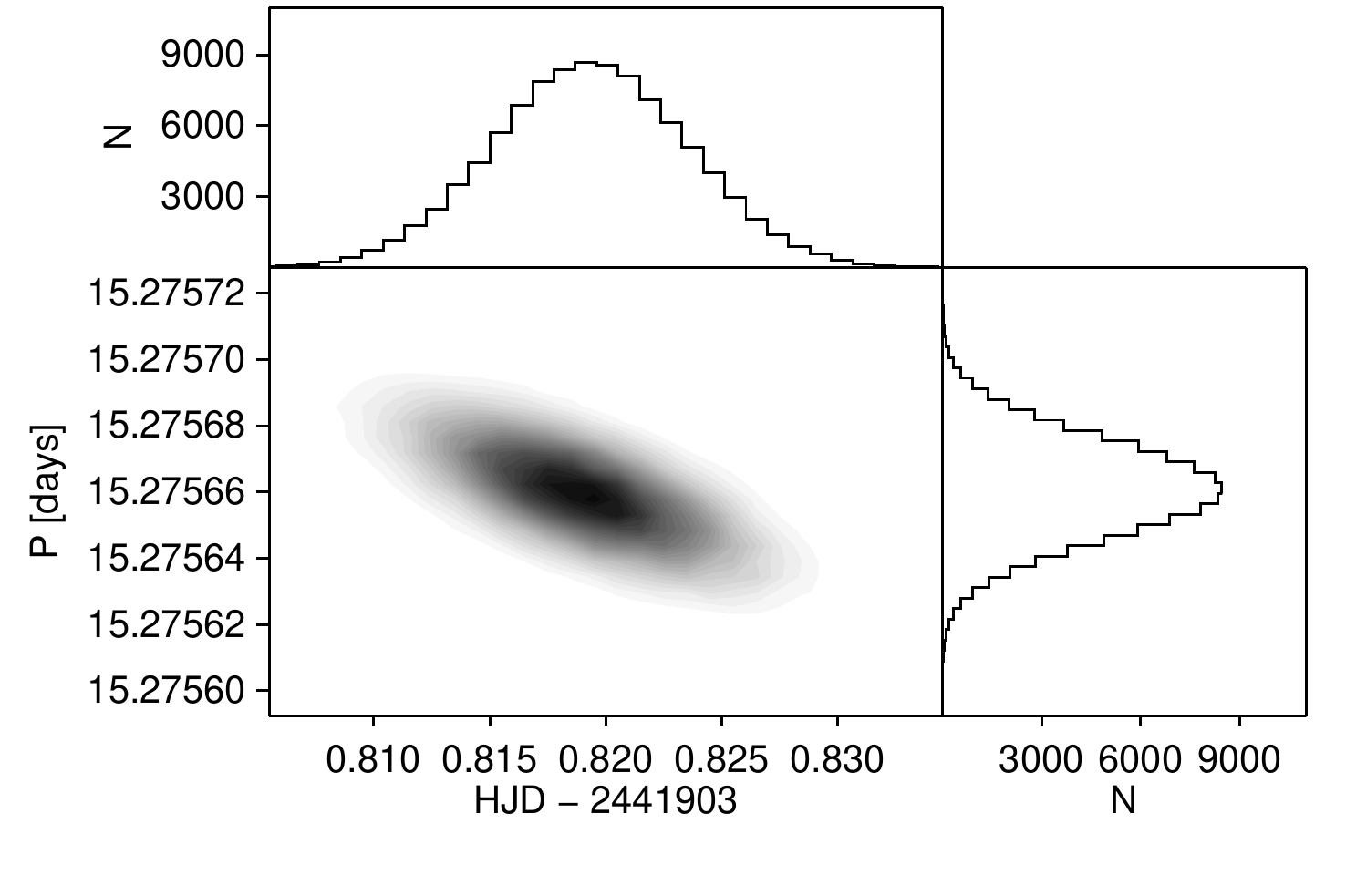}
   \caption {\label{fig:times} {\bf Results for the timing
       parameters}, based on our Monte Carlo analysis of the times of
     minimum light given by \citet{ahn1992}. The gray shading indicates
     the density of Monte Carlo results. The marginalized probability
     distributions are shown on the sides.}
  \end{center}
\end{figure}

Our spectral model is similar to the models described in Papers~I and
II. Because of the light from the foreground star, we cannot treat the
RM effect as a simple wavelength shift, as is commonly done for
eclipses of stars by planets. Instead we modeled the line profiles of
both stars simultaneously, taking into account stellar rotation,
surface velocity fields, orbital motion, and partial blockage during
eclipses. By adjusting the parameters of the model to fit the observed
spectra, we derived estimates of the orbital and stellar parameters.

For each star, and for each phase of the eclipse, we created a
discretized stellar disk with about 200,000 pixels in a cartesian
  coordinate system. We assumed the disk to be circular, since the
  stars are well detached and have slow rotation speeds relative to
  the breakup velocity. Each pixel has its own emergent spectrum,
weighted in intensity according to a linear limb-darkening law, and
Doppler shifted due the combined effects of orbital motion, rotation,
and macroturbulence. The orbital motion is specified by a Keplerian
model common to all pixels. The rotation is assumed to be uniform (no
differential rotation). Following \citet{gray2005}, the macroturbulent
velocity field is assumed to obey Gaussian distributions for the
tangential and radial components, with equal amplitudes, brightnesses,
and surface fractions.

\begin{figure*} [t]
\begin{center}
\includegraphics[width=16.5cm]{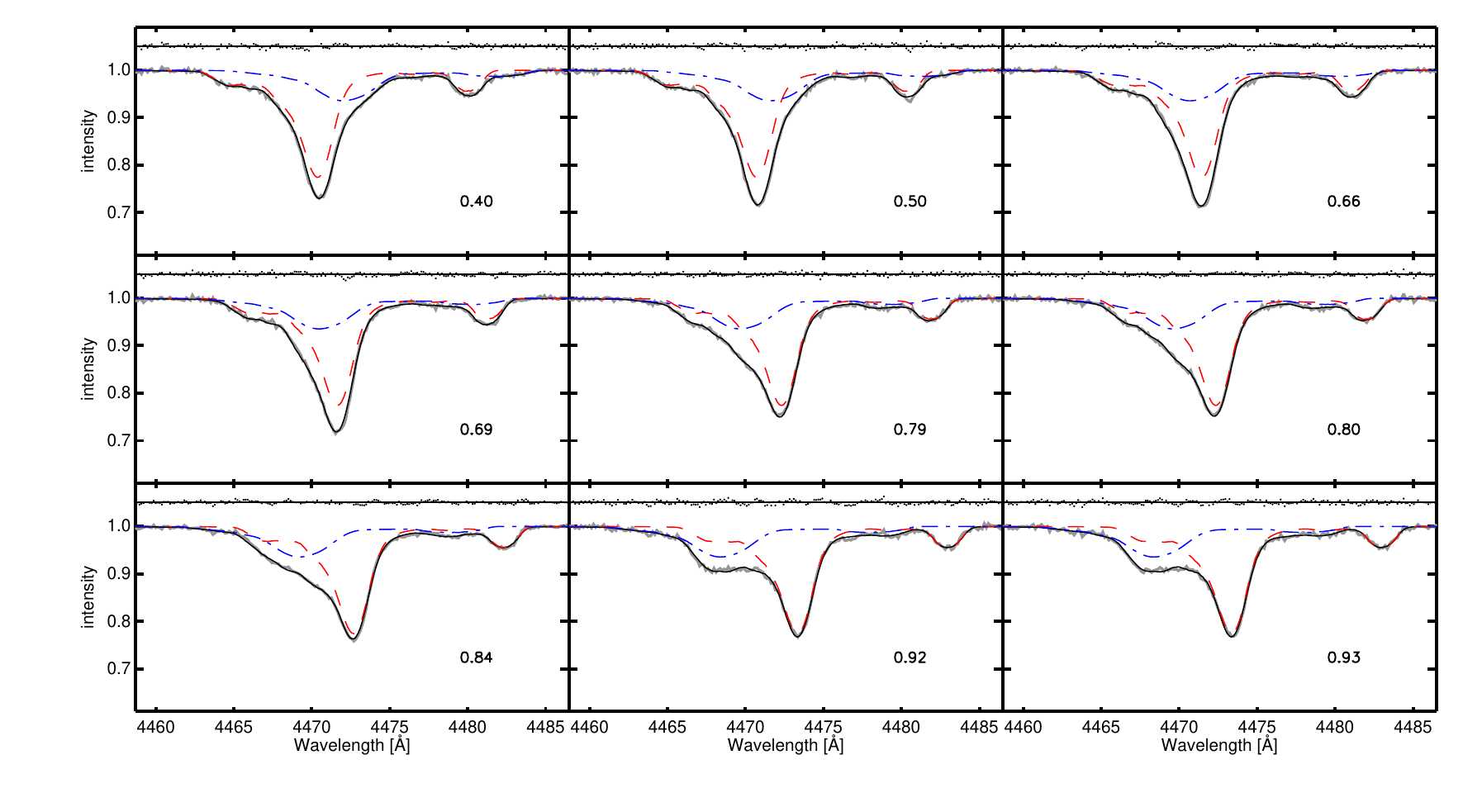}
\caption {\label{fig:shape_out_of_eclipse} {\bf Observations of
    NY\,Cep outside of eclipses.} These panels show spectra of both
  stars in the NY\,Cep system in the spectral region around the He\,I
  line at 4471\,\AA{} and Mg\,II line at 4481\,\AA{}. The number in
  each panel indicates NY\,Cep's orbital phase at time of observation,
  defined such that phase 0 corresponds to periastron. The gray
  solid lines represent the data and the (red) dashed and (blue)
  dash-dotted lines are the simulated absorption lines of the primary
  and secondary, respectively. The black line is the best fitting
  model. The dots around the line at a flux level of 1.05 represent
  the differences between the data and the model.}
\end{center}
\end{figure*}

The sky-plane coordinates of the foreground and background stars are
calculated, and if the background star is being eclipsed then the
eclipsed pixels are assigned zero intensity.  The emergent spectra
from the uncovered pixels of both stars are summed to create a model
absorption line kernel. For the He\,I line a further convolution with
a Lorentzian function is applied, to account for pressure
broadening. Figure~\ref{fig:model_example} illustrates the effects of
rotation, macroturbulence, pressure broadening, and eclipse blockage
on the model absorption line kernel.

The kernel is then convolved with a line-list in the wavelength region
between 4460\,\AA{} and 4485\,\AA.{} The positions and line strengths
are obtained from the Vienna Atomic Line Database (VALD)\footnote{{\tt
    http://ams.astro.univie.ac.at/vald/}} for the stellar parameters
given by \cite{holmgren1990} and are listed in Table~\ref{tab:nycep}.
The He\,I, Mg\,II, Al\,III and some of the O\,II lines consist of
multiple lines with energy levels spaced closely enough ($\le
0.3$\,\AA{}) that we model them as single lines. We also omitted one
S\,III line which has nearly the same wavelength as one of the O\,II
lines.

\subsection{Eclipse timing}
\label{sect:eclipse_timing}

Since we did not obtain new photometric data, we relied on the data
presented by \citet{ahn1992} (in their Table 2) to constrain the
orbital period and time of primary minimum light. The lack of error
estimates by \citet{ahn1992} presented a complication, which we dealt
with as follows. We fitted a linear function of epoch to the reported
times, assuming equal errors in all the measurements, determined by
the requirement $\chi^2 = N_{\rm dof}$.  The resulting error in each
time was 0.012~days, and the resulting ephemeris are given in
Table~\ref{tab:times}. Our result for the period agrees with that
given by \citet{ahn1992}, although we found a slightly different time
of primary minimum light, presumably due to different weighting of the
measurements.

\begin{table}[h]
  \begin{center}
    \caption{Results eclipse timings\label{tab:times}}
    \smallskip
    \smallskip
    \smallskip
    \begin{tabular}{l r@{$\pm$}l l}
      \tableline\tableline
      Parameter  &  \multicolumn{2}{c}{This work} & \cite{ahn1992}\\
      \tableline
      Period & 15\fd27566&0.00002  & 15\fd27566 \\
      T$_{\rm min, I}$~1973 [HJD-2\,400\,000] &  41903.819 & 0.006  & 41903.8161\\
      T$_{\rm min, I}$~2009 [HJD-2\,400\,000] &  55117.265 & 0.016  & \\
      \tableline
    \end{tabular}
  \end{center}
\end{table}

Then we used a Monte Carlo approach to calculate the predicted times
of minimum light at the epochs of our observations, in a manner that
respects the correlation between the parameters of the ephemeris. We
created $10^5$ fake data sets by adding Gaussian perturbations to the
times of minima of \citet{ahn1992}, with a standard deviation of
0.012~days. We then fitted a line to each of these fake data sets and
used the linear fit to calculate the expected mideclipse time on
2009~Oct.~12/13, the last night of our eclipse observations. The mean
and standard deviation among all $10^5$ results were taken to be the
value and the error in the mideclipse time.  The results are given in
Table~\ref{tab:times}.

\subsection{The fitting procedure}
\label{sect:fitting_procedure}

\begin{figure*}[t]
 \begin{center}
  \includegraphics[width=16.5cm]{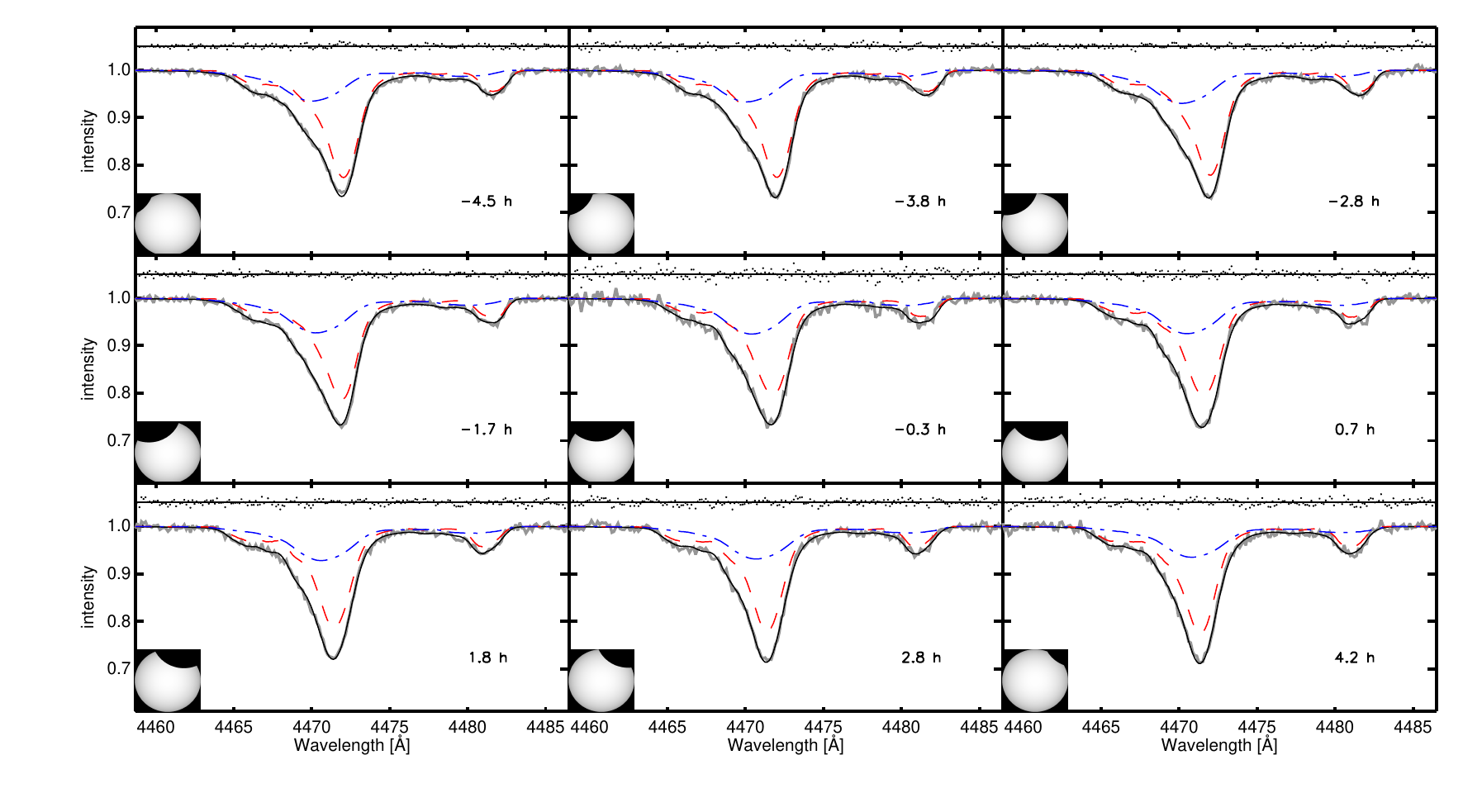}
  \caption {\label{fig:shape_primary_eclipse} {\bf Observations of
      NY\,Cep during primary eclipse.}  Similar to Figure~\ref{fig:shape_out_of_eclipse},
    but for spectra obtained during primary eclipse. The number in
    each panel indicates the time from mideclipse, in hours.  The
    corner of each panel shows an illustration of the eclipse
    phase. In the best-fitting model, $\beta_{\rm p} = 2\pm4^{\circ}$,
    indicating a close spin-orbit alignment.  Focusing on the Mg\,II
    line, one can see in particular for the observations at $-$2.8\,h
    and 2.8\,h how the RM-effect creates a redshift during the first
    half of the eclipse and a blue shift during the second half of the
    eclipse.  For the observation at $-$0.3\,h, near mideclipse, blue
    and redshifted light are blocked by similar amounts.}
 \end{center}
\end{figure*}

We now describe the model parameters in detail. Seven parameters
describe the Keplerian orbit: the time of minimum light ($T_{\rm
  min, I}$), the eccentricity ($e$), the argument of periastron
($\omega$), the velocity semiamplitudes of the primary and secondary
(K$_{\rm p}$ and K$_{\rm s}$), and the velocity offsets ($\gamma_{\rm
  p}$ and $\gamma_{\rm s}$).\footnote{The velocity offsets are
  approximately equal to the radial velocity of the NY\,Cep barycenter
  with respect to the Solar system barycenter, but the offsets can
  differ from each other due to subtle factors specific to each star,
  such as the gravitational redshift and line blending.  The latter
  term refers to the fact that the stars have slightly different
  strengths and positions of closely spaced lines (unresolved in our
  spectra), due to the differences in effective temperature, surface
  gravity, and other atmospheric parameters. These differences can
  result in a systematic bias in the inferred systemic velocity.} We
held the period ($P$) fixed, as its uncertainty is negligible over the
3 months spanned by our observations.

Another 4 free parameters specify the photometric aspects of the
eclipse: the light ratio  between the two stars at the wavelength
of interest (L$_{\rm s}$/L$_{\rm p}$ at 4500\,\AA{}), the fractional
radii of the stars ($r_{\rm p}$ and $r_{\rm s}$), and the orbital
inclination ($i_o$). The linear limb darkening parameter (u$_{\rm i}$)
was held fixed at $0.4$ for both stars \citep{gray2005}.

The model of the velocity fields introduces 6 parameters: the
projected equatorial rotation speeds ($v\sin i_{\rm p}$ and $v\sin
i_{\rm s}$), the Gaussian width of the macroturbulence for the primary
star ($\zeta_{RT}P$), the half-width at half-maximum of the Lorentizan
function representing pressure broadening for each star ($\xi_{\rm p}$
and $\xi_{\rm s}$), and finally, the parameters of greatest interest
for this study, the sky-projected spin-orbit angle ($\beta_{\rm p}$).
The angle is defined according to the convention of
\citet{hosokawa1953}, such that $\beta=0^{\circ}$ when the axes are
parallel, $\beta=\pm 90^{\circ}$ when they are perpendicular, and
$\beta>0$ when the time-integrated RM effect is a redshift for an
 orbit with $i_{o}<90^{\circ}$.\footnote{The parameter $\lambda$, which was introduced
  by \cite{otha2005} and is commonly used in the exoplanet community,
  is simply $-\beta$.}  The macroturbulence parameter for the
secondary was held fixed to 20\,km\,s$^{-1}$, since the secondary is
significantly fainter than the primary, causing the results to be
insensitive to this parameter.\footnote{Indeed, although we included
  macroturbulence for completeness, the key results for $\beta_{\rm
    p}$ do not depend on the details, and similar results were
  obtained even when macroturbulence was ignored.  One can see from
  Figure~\ref{fig:model_example} that macroturbulence has only a small
  effect on the overall shape of the model absorption line, given the
  high $v \sin i$ of the primary. This is even more true of the
  secondary, which rotates about twice as fast as the primary.}

Eight additional parameters are needed for each star to describe the
relative depth of the spectral lines in the wavelength range being
modeled. In addition, 3 free parameters are needed to describe the
quadratic function used to normalize the continuum level of each of
the 46 spectra.  The total number of parameters is therefore
7~(orbital)~$+$ 4~(photometric)~$+$ 6~(velocity fields)~$+$ $8\times
2$~(line depths)~$+$ $3\times 46$~(normalization), for a total of 171
adjustable parameters. Although this may seem like a large number, the
138 normalization parameters are ``trivial'' in the sense that they
can be optimized separately in subsets of 3, for a given choice of the
other parameters. Also, several of the other parameters are subject to
{\it a priori} constraints. More detail on these points is given
below.

The fitting statistic was
\begin{eqnarray}
\chi^{2} & = & \sum_{i=1}^{46}  \sum_{j=1}^{213}
\left[\frac{f_{\lambda_{i,j}}({\rm obs})-f_{\lambda_{i,j}}({\rm calc})}{\sigma_{i}}\right]^2+\;\;\;\nonumber\\
&  &   \left(\frac{{\rm T}_{\rm min  I} - 2455117.265}{0.016}\right)^2 + \;\;\;\nonumber\\
&  & \left(\frac{i_{o} -78^{\circ}}{1^{\circ}}\right)^2 + \left(\frac{{\rm  L}_{{\rm s}}/{\rm  L}_{{\rm p}} - 0.37}{0.06}\right)^2 + \;\;\;\nonumber\\
&  &  \left(\frac{R_{\rm p}/R_\odot - 6.8}{0.7}\right)^2 +
      \left(\frac{R_{\rm s}/R_\odot - 5.4}{0.5}\right)^2, \nonumber
\end{eqnarray}
where $f_{\lambda_{i,j}}({\rm obs})$ is the observed flux in pixel $j$
in observation $i$ and $f_{\lambda_{i,j}}({\rm calc})$ is the
calculated flux based on a particular choice of the model parameters,
and ${\sigma_{i}}$ represents the uncertainty in the flux pixel for
observation $i$. The latter was taken to be the reciprocal of the SNR,
estimated as described in section~\ref{sect:observations}. In this
equation, the first two terms represent the usual sum-of-squares, and
the remaining terms represent Gaussian priors based on the results of
\citet{holmgren1990} (except for T$_{\rm min, I}$, which was discussed
in the previous section).

To optimize the parameters we used Levenberg-Marquardt (LM)
least-squares minimization. The main optimization was conducted for
the nontrivial parameters. However, whenever $\chi^2$ was computed
within that process, the values of the trivial (normalization)
parameters were first optimized using a separate 3-parameter
minimization for each order. This process is equivalent to the
``Hyperplane Least Squares'' method that was described and tested by
\citet{bakos2010}. To estimate the parameter uncertainties, we used
the bootstrap method described by \citet{press1992}, with
5$\times$10$^3$ realizations.

With 171 parameters, 5 Gaussian priors, and $46\times 213$ data
pixels, there are 9632 effective degrees of freedom.  The best-fitting
model has $\chi^2 = 9148$, or ${\overline{\chi}}^2 = 0.95$. The low
${\overline{\chi}}^2$ is probably the result of underestimating the
SNR of the relevant lines. The estimate was based on the continuum
nearer to the edges of the order, where the SNR is lower (see
Figure~\ref{fig:nycep_spec}). We have not attempted to correct for
this effect.

\begin{figure} [t]
 \begin{center}
  \includegraphics[width=8.5cm]{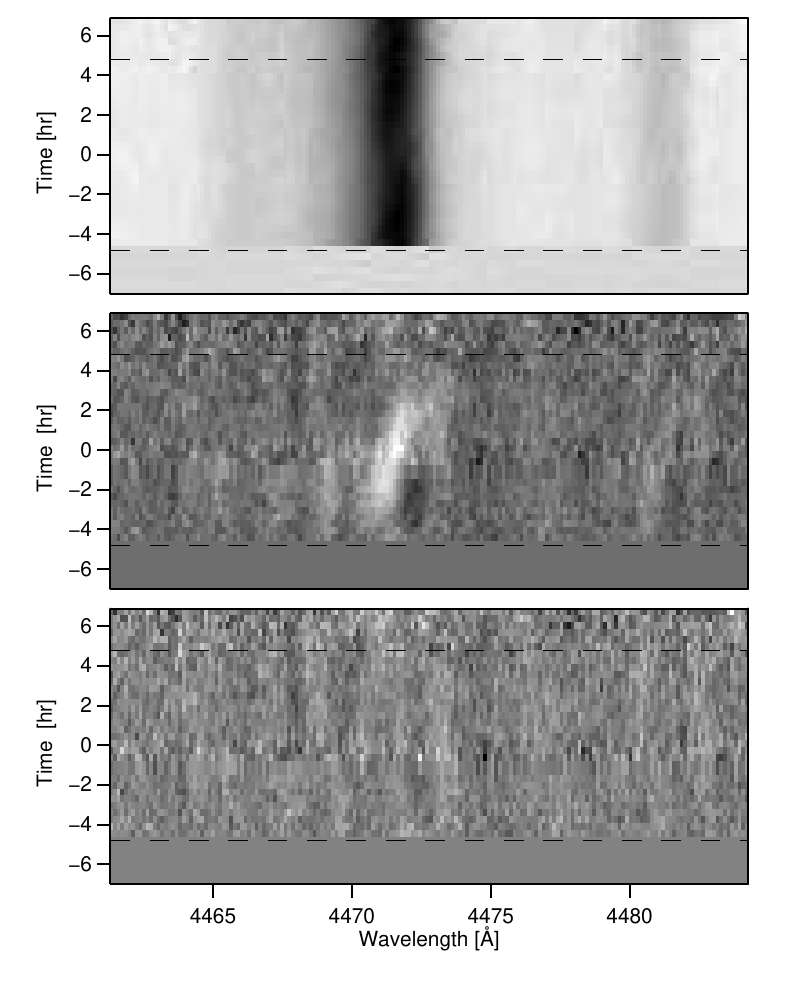}
  \caption {\label{fig:tv_primary_eclipse} {\bf Observations of
      NY\,Cep during primary eclipse.} {\it Top.}---Grayscale
    depiction of the spectra obtained during and after primary
    eclipse. Each spectrum is shifted into the rest frame of the
    primary star, and the best-fitting model of the secondary spectrum
    has been subtracted. The most prominent lines are He\,I (seen here
    blended with the weaker O\,II lines) and Mg\,II  4481\,\AA\,.
    Dashed lines indicate the beginning and end of the eclipse. {\it
      Middle.} ---The same as the top panel, after subtracting the mean
    absorption line profile shown in Figure~\ref{fig:lines}.  The RM
    effect is evident as the white residual traveling from
    the blue to the red as the eclipse progresses. {\it Bottom.}---
    Residuals between best fitting model and data.}
 \end{center}
\end{figure}

\section{Results}
\label{sect:results}

\begin{figure} [t]
  \begin{center}
    \includegraphics[width=8.5cm]{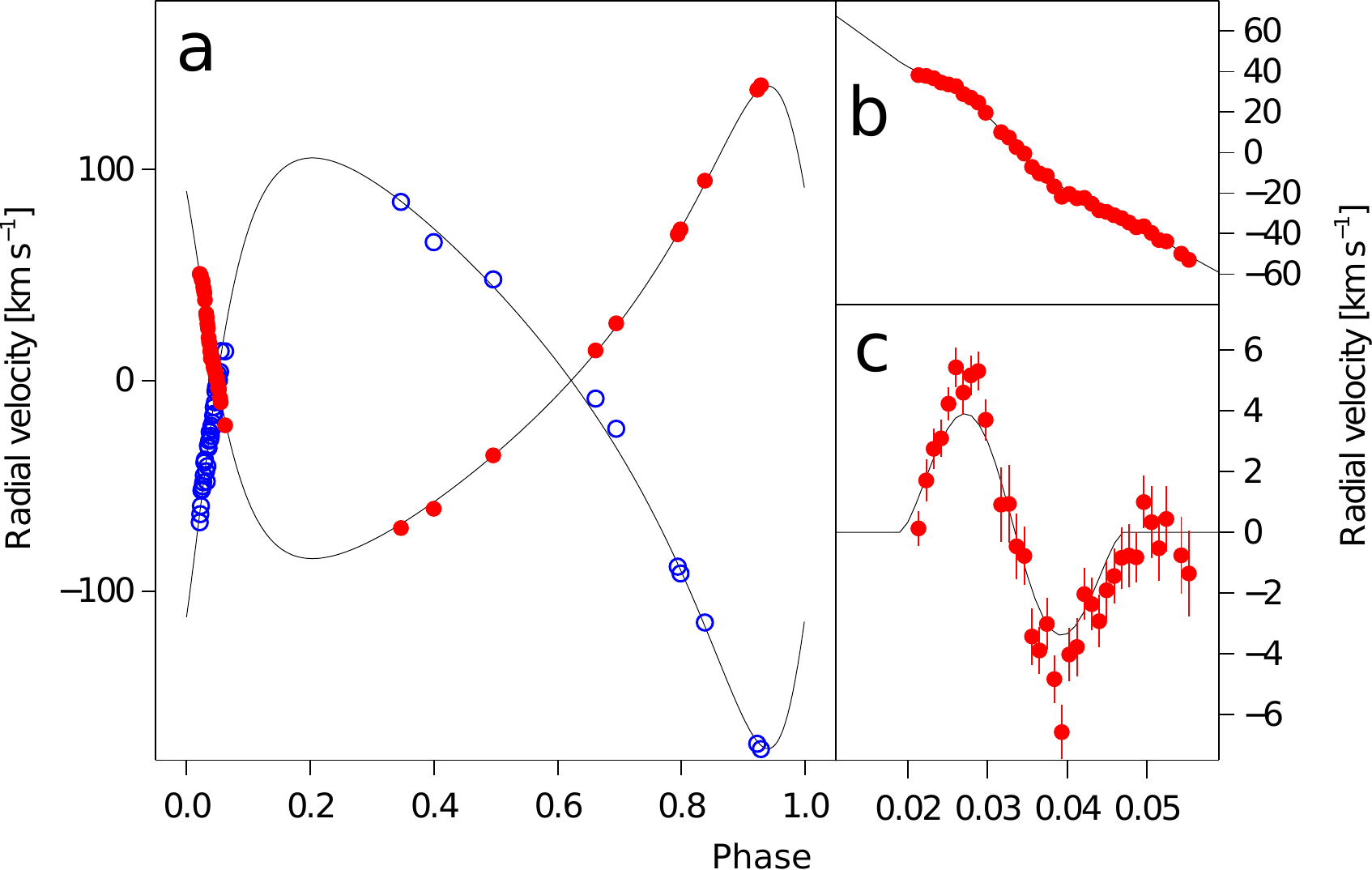}
    \caption {\label{fig:rv} {\bf Apparent radial velocities of NY\,Cep.} 
      {\bf a:} The apparent radial velocity (RV) of the primary (red filled
      circles) and secondary (blue open circles) as a function of orbital
      phase. The line positions were determined by fitting Gaussian
      functions. The solid line is the calculated radial velocity based on
      our model to the shape ob the absorption lines, including
      Keplerian motion and the RM effect. {\bf b,c:}  Close-ups of the
      RM effect during the primary eclipse. The upper
      panel shows the RM effect plus orbital motion and the lower panel
      shows only the RM effect, a redshift during the first half
      of the eclipse, and a blueshift during the second half. One can see
      that a fit to the RV data would overestimate the amplitude of  RM signal.
      We emphasize that the eclipse radial velocities are shown here for
      illustration only, as a concise visual summary of the complex
      distortions of the eclipse spectra: they were not used in our
      quantitative calculations.}
  \end{center}
\end{figure}

The results for the model parameters are given in
Table~\ref{tab:results}.  Figure~\ref{fig:shape_out_of_eclipse} shows
the model fitted to a few out-of-eclipse spectra in the vicinity of
the He\,I and Mg\,II lines.  Figure~\ref{fig:shape_primary_eclipse}
shows the same for a few of the spectra obtained during primary
eclipse. A grayscale representation of the eclipse spectra and
post-egress spectra is shown in figure~\ref{fig:tv_primary_eclipse}.
We also display the apparent radial velocities in NY\,Cep in
figure~\ref{fig:rv}. 

\subsection{Orbital parameters}
\label{sect:orbital}

Our result for T$_{\rm min I}$ is 1.2$\sigma$ away from the value
calculated from the older eclipse timings.  Since the ephemeris is
based on observations obtained more than 20 years ago, and we do not
know the true uncertainty in each measurement, we regard this as good
agreement. Likewise the eccentricity is within 2$\sigma$ of the value
given by \citet{holmgren1990}. The measured value for $\omega$ agrees
with the previous value, but as it is expected to change over time due
to apsidal motion a more careful comparison is needed, as discussed in
section~\ref{sect:discussion}.

Our results for K$_{\rm p}$ and $\gamma_{\rm p}$ agree with the values
found by \cite{holmgren1990}. However, our results for the secondary
star are different: we find a lower value for K$_{\rm s}$, and we
found $\Delta\gamma \equiv \gamma_{\rm s} - \gamma_{\rm p}$ to be
consistent with zero. In contrast, \cite{holmgren1990} found the
secondary to be redshifted by $26$\,km\,s$^{-1}$ relative to the
primary, and \cite{heard1968} found an even greater difference of
$32$\,km\,s$^{-1}$. We speculate that the previous measurements were
subject to bias because of the relative faintness of the secondary,
its rapid rotation, and the need to use pressure-broadened (and
asymmetric) lines. Based on the agreement in $\gamma_{\rm p}$ it
seems that the systemic velocity has not changed significantly
over the last decades.

\begin{table*}[t]
 \begin{center}
  \caption{Parameters of the NY\,Cephei system.\label{tab:results}}
    \smallskip 
       \begin{tabular}{l  r@{$\pm$}l r@{$\pm$}l   }
          \tableline\tableline
	  \noalign{\smallskip}
	  Parameter &  \multicolumn{2}{c}{ This work}  &    \multicolumn{2}{c}{Literature values} \\
	  \noalign{\smallskip}	 
	  \hline
	  \noalign{\smallskip}
          \multicolumn{5}{c}{Orbital parameters} \\
          \noalign{\smallskip}
          \hline
	  \noalign{\smallskip}
          Time of primary minimum, T$_{\rm min, I}$ [HJD$-$2\,400\,000]         &   55117.287 & 0.009                                 & \multicolumn{2}{l}{41903.8161\tablenotemark{$\dagger$}}      \\
          Period, $P$ [days]                                                                &     \multicolumn{2}{c}{15.27566 (fixed)}   &   \multicolumn{2}{l}{15.27566\tablenotemark{$\dagger$} }   \\
          Eccentricity, $e$                                                                &    0.443 & 0.005                                          &      0.48 & 0.02\tablenotemark{$\star$}       \\
          Argument of periastron, $\omega$ [deg]                    &     56.3 & 1                                               &      58 & 2\tablenotemark{$\star$}        \\   
          Orbital inclination, $i_{o}$ [deg]                                &   78.8 & 0.7               &   78&1\tablenotemark{$\star$}   \\   
          Velocity semiamplitude (primary), K$_{\rm p}$ [km\,s$^{-1}$]              &     113.8 & 1.2                                        &      112 & 2\tablenotemark{$\star$}          \\
          Velocity semiamplitude (secondary), K$_{\rm s}$ [km\,s$^{-1}$]              &        139 & 4                                   &      158 & 8\tablenotemark{$\star$}       \\
          Velocity offset (primary), $\gamma_{\rm p}$ [km\,s$^{-1}$]   &     $-17.3$ & 0.7                                   &     $-15$  & 2\tablenotemark{$\star$}             \\
          Velocity offset (secondary), $\gamma_{\rm s}$ [km\,s$^{-1}$]   &   $-21$ & 4                                      &   $9$  & 6\tablenotemark{$\star$}         \\
          \noalign{\smallskip}
          \hline
          \noalign{\smallskip}
          \multicolumn{5}{c}{Stellar  parameters} \\
          \noalign{\smallskip}
          \hline 
          \noalign{\smallskip}
          Light ratio, L$_{\rm s}$/L$_{\rm p}$                      &     0.36 & 0.05                                              & 0.37&0.06\tablenotemark{$\star$}       \\
          Fractional radius (primary), r$_{\rm p}$                              &          0.086  & 0.007                       &       \multicolumn{2}{c}{}      \\
          Fractional radius (secondary), r$_{\rm s}$                               &      0.084 &  0.009                   &      \multicolumn{2}{c}{}        \\
          Projected rotation speed (primary),	  $v \sin i_{\rm p}$ [km\,s$^{-1}$]     &      78 & 3                                                  & 75 & 10\tablenotemark{$\star$}  \\
          Projected rotation speed (secondary),	  $v \sin i_{\rm s}$ [km\,s$^{-1}$]     &      155 & 6                                        & 125&14\tablenotemark{$\star$}       \\
          Macroturbulence parameter (primary),  $\zeta_{\rm p}$ [km\,s$^{-1}$]    &    23 & 6                              & \multicolumn{2}{c}{}          \\
          Macroturbulence parameter (secondary),  $\zeta_{\rm s}$ [km\,s$^{-1}$]    &  \multicolumn{2}{c}{20 (fixed)}                       & \multicolumn{2}{c}{}             \\
          Pressure broadening parameter (primary),        $\xi_{\rm p}$ [km\,s$^{-1}$]      &    36 & 1                                                   & \multicolumn{2}{c}{}          \\
          Pressure broadening parameter (secondary),   $\xi_{\rm s}$ [km\,s$^{-1}$]      &    54 & 6                                                   & \multicolumn{2}{c}{}             \\
          Linear limb darkening parameter (primary),     u$_{\rm p} $                                    &  \multicolumn{2}{c}{0.4 (fixed)}                     &     \multicolumn{2}{c}{}              \\
          Linear limb darkening parameter (secondary),   u$_{\rm s} $                                    &   \multicolumn{2}{c}{0.4 (fixed)}                      &     \multicolumn{2}{c}{}              \\	
          Projected spin-orbit angle (primary),   $\beta_{\rm p}$ [$^{\circ}$]             &      2 & 4                                                  & \multicolumn{2}{c}{}       \\ 
          \noalign{\smallskip}
          \hline
          \noalign{\smallskip}
          \multicolumn{5}{c}{ Indirectly derived parameters} \\
	  \noalign{\smallskip}
          \hline
          \noalign{\smallskip}
           Projected orbital semimajor axis,    $a \sin i$ [R$_{\odot}$]                &       68.6&1                                                         &     71.4&2.2\tablenotemark{$\star$}       \\
	  $M_{\rm p} \sin^{3}i$ [$M_{\odot}$]   &     \multicolumn{2}{c}{10.1$^{+0.8}_{-0.6}$}                              &   12    & 1\tablenotemark{$\star$}             \\
	  $M_{\rm s} \sin^{3}i$ [$M_{\odot}$]   &     \multicolumn{2}{c}{8.3$^{+0.5}_{-0.3}$}                                &  8.7  & 0.6\tablenotemark{$\star$}                 \\
          Primary mass,          $M_{\rm p} $ [$M_{\odot}$]               &     \multicolumn{2}{c}{10.7$^{+0.9}_{-0.7}$}            &     13  & 1\tablenotemark{$\star$}                 \\
          Secondary mass,	  $M_{\rm s} $ [$M_{\odot}$]               &    \multicolumn{2}{c}{8.8$^{+0.6}_{-0.4}$}              &      9 & 1\tablenotemark{$\star$}                   \\
          Primary radius,         $R_{\rm p} $ [$R_{\odot}$]               &     6.0 & 0.5                  &     6.8  & 0.7\tablenotemark{$\star$}                 \\
          Secondary radius,	  $R_{\rm s} $ [$R_{\odot}$]               &  5.8 &  0.5                     &      5.4 & 0.5\tablenotemark{$\star$}                   \\
	  \noalign{\smallskip}
	  \tableline

           \noalign{\smallskip}
           \noalign{\smallskip}
            $\dagger$  Data from \cite{ahn1992}\\
            $\star$  Data from \cite{holmgren1990}
        \end{tabular}
     \end{center}
  \end{table*}

\subsection{Stellar parameters}

We further find with our fit to all spectra a luminosity ratio
(L$_{\rm s}$/L$_{\rm p}$) of $0.36\pm0.05$ at $4500$\AA{}. This result
agrees with the value $0.35$ that is obtained from comparing the
equivalent widths of the He\,I (4471\,\AA{}) line profiles shown in
Figure\,\ref{fig:lines}.\footnote{In estimating the luminosity
  ratio from the EWs, we accounted for the small increase
  ($\sim$10$\%$) in the EW of this line towards lower temperatures in
  the spectral range from to B0 to B2. See Figure\,1 of
  \cite{leone1998}.} Both values are consistent with $0.37\pm0.06$
found by \cite{holmgren1990}, based on spectrophotometry.

Our model gives projected rotation speeds of $78\pm1$ km\,s$^{-1}$ for
the primary $155\pm4$ km\,s$^{-1}$ for the secondary. The width of the
Gaussian function describing the macroturbulence in the primary is
$23\pm3$ km\,s$^{-1}$, and the half-width at half-maximum of the
Lorentzians describing pressure broadening are $36\pm1$ km\,s$^{-1}$
and $54\pm6$ km\,s$^{-1}$. These error estimates must be understood as
internal to our model, which we recognize may not be completely
realistic, especially as it pertains to pressure broadening. We have
assumed that absorption lines form at a specific value of pressure,
which is not necessarily true. For this reason, the Lorentzian widths
have no simple physical interpretation, and the results may be
somewhat biased for all other parameters affecting line broadening
(namely $v\sin i$ and $\zeta$). For comparisons to other analyses
obtained with different instruments and different analysis procedures,
we recommend using the more conservative error estimates given in
Table~\ref{tab:results}. For the projected rotation rates, we find a
similar result for the primary as did \cite{holmgren1990}, but we find
a significantly higher $v\sin i$ for the secondary, perhaps for the
same reasons given in section \ref{sect:orbital}.

\begin{figure} [t]
 \begin{center}
  \includegraphics[width=7.5cm]{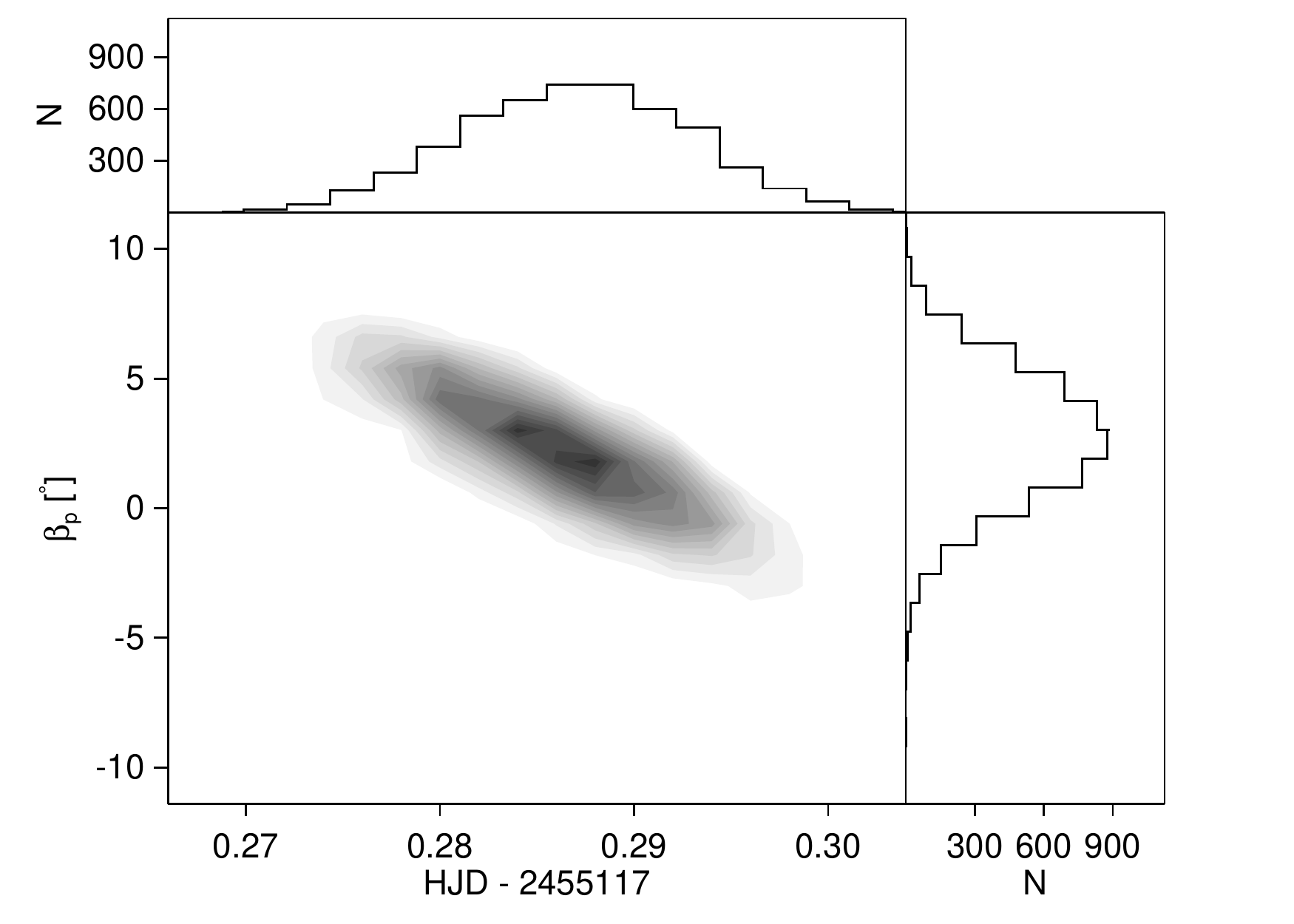}
  \caption {\label{fig:corrs} {\bf Results for time of minimum light
      and} $ \mbox{\boldmath$ \beta_{\rm p} $}$, based on our analysis
    of the stellar spectra using the bootstrap method. There is a
    strong correlation between these two parameters.}
 \end{center}
\end{figure}

As for the projected spin-orbit angle $\beta_{\mathrm{p}}$, the focus
of this study, we find $\beta_{\mathrm{p}}=2\pm4^{\circ}$. There is no
correlation between this parameter and the linewidth parameters, so
the concerns raised above about the oversimplified pressure-broadening
model do not apply here. A strong correlation does exist with the time
of minimum light (see Figure~\ref{fig:corrs}). This implies that
future photometric observations to refine the eclipse ephemeris would
lead to smaller uncertainty in $\beta_{\rm p}$, although the
uncertainty is already quite small.

The absolute radii of the stars can be calculated from the fractional
radii and the other orbital parameters. We find radii consistent with
the values given in the literature, $R_{\rm
  p}$~=~6.0$\pm$0.5\,$R_{\odot}$ and $R_{\rm
  s}$~=~5.8$\pm$0.6\,$R_{\odot}$, which is not surprising since our
results were strongly influenced by the priors on those parameters.
For the stellar masses, we find $M_{\rm
  p}$~=~10.7$\pm$0.8\,$M_{\odot}$ and $M_{\rm
  s}$~=~8.8$\pm$0.5\,$M_{\odot}$ for the primary and secondary,
respectively. The value found for the primary is smaller than that
reported previously, due to the previously mentioned discrepancy in
the velocity semiamplitude of the secondary.

\section{Discussion}
\label{sect:discussion}

\paragraph{Primary spin axis} 
We have measured the angle between the projections of the orbital and
stellar angular momentum vectors for the primary star in NY\,Cep and
find that these projections are aligned within the uncertainty of our
measurement ($\beta_{\rm{p}}=2\pm4$$^{\circ}$). We believe this result
to be robust, even if our model is simplified in many respects.  We
have neglected any changes in the limb-darkening law within the lines
as compared to the continuum, as well as a number of other
``second-order'' effects, such as gravity brightening and differential
rotation.  We experimented with more complex models including these
phenomena and found that they produced effects too small to affect our
conclusions about spin-orbit alignment. Likewise we experimented with
a ``two-layer'' model for the pressure-broadened lines [see
figure~11.5 of \cite{gray2005}] and found that while it provided a
slightly better fit to the data, it introduced several new free
parameters and led to no significant changes in any other parameters.

Our result for $\beta_{\rm p}$ gives a lower bound on the true angle
between the stellar and orbital spins, the obliquity ($\psi$). We
however expect the obliquity to be not much larger than $\beta$ unless
the primary spin axis is highly inclined towards the line of sight, an
unlikely scenario.

\paragraph{Absolute dimensions}
For our discussion we adopt the effective temperatures derived by
\cite{holmgren1990}, which are printed in
table~\ref{tab:nycep}. Because we do not have any new photometry, we
do not attempt to derive new effective temperatures or a new age for
the system. We adopt the previous classification of NY\,Cep as a young
system ($\sim$$6$ Myr) at a distance of $750\pm90$\,pc, making it
consistent with membership in the Cepheus OB~III association.  Our
finding of a lower mass for the primary does not change the picture of
NY\,Cep significantly given the uncertainties in other parameters.

\paragraph{Stellar rotation}
If the spin periods were equal to the orbital period, then the stellar
rotation velocities would be 20 km\,s$^{-1}$ and 19 km\,s$^{-1}$, for
the primary and secondary, respectively. These numbers are smaller
than our results for $v\sin i$, which are $78\pm3$km\,s$^{-1}$ and
$154\pm6$ km\,s$^{-1}$. Assuming $\psi=0$ for both stars, the implied
rotation periods are 3.9 and 1.9 days.

The stars do not seem to be pseudosynchronized, either.  The ratio of
rotational to orbital frequencies is 3.9 and 8.0 for the
primary and secondary, respectively.  In the \citet{hut1981} model of
pseudosynchronization, the ratio would be 2.3, which is lower than the
observed ratios.  If the stars had been significantly influenced by
tidal evolution, then one would expect the rotational frequency to be
smaller for the secondary star than for the primary star, because the
less massive star should synchronize first.  In contradiction with
this expectation, we observe the secondary to have a higher rotation
frequency than the primary.

All of these findings are in agreement with a picture in which the
system is only a few million years old, and tidal forces have not had
enough time to alter the rotational state of the stars after their
arrival on the zero-age main sequence.

\paragraph{Apsidal motion}
With the data at hand it is not possible to investigate the internal
structures of the two stars in a stringent way via observations of
apsidal motion. Firstly, the uncertainties in the radii and ages are
so high that the theoretical expected apsidal motion has a high
uncertainty. Using the system parameters, together with internal structure
constants of $k_{2} = -0.080\pm0.35$\footnote{We estimated the
  apsidal motion constants using tables from \cite{claret1995}.} for
both stars, we derive an expected apsidal motion rate of
$19^{+41}_{-9}$ arcsec per cycle. Secondly, the measurements of the
argument of the periastron ($\omega$) are also quite uncertain. Using
the values from \cite{heard1968}, who derived an $\omega$ of
$50.9\pm3.7$, from \cite{holmgren1990}, and from this work, we derive
a measured apsidal motion rate of $6\pm11$ arcsec per cycle. Therefore
the expected and measured apsidal motions are consistent, but the
bounds are too weak to give additional information. To make progress
on this field both the expected and measured values for the apsidal
motion have to have greater precision and accuracy.

\section{Conclusion}
\label{sect:conclusion}

Through an analysis of high resolution spectra of the NY\,Cep system,
observed during and outside of primary eclipse, we have calculated the
orbital and primary and secondary parameters. We also reanalyzed times
of minima from the literature to obtain an updated ephemeris for the
epoch of our spectroscopic observations. We find a $\sim$15\% smaller
mass for the primary star than earlier studies, and derive results for
the other parameters that are similar to those reported in the
literature values.

We measured the angle on the sky between the primary spin axis and the
orbital axis by exploiting the Rossiter-McLaughlin (RM) effect. We
find that the projections of these axes on the sky are closely aligned
($\beta_{\rm p}=2\pm4$). The close alignment does not seem to have
been the result of tidal evolution, because tidal evolution would also
have synchronized the rotational and orbital periods, which is not
observed.

The finding of a close spin-orbit alignment is in strong contrast to
the situation in the other young, early type, detached close binary,
DI\,Her, for which the orientation of the axes has been measured
(Paper II). For that system we found strongly misaligned spin and
orbital axes. While NY\,Cep and DI\,Her are similar in that they both
harbor young, well detached, early type stars, and have high orbital
eccentricities their formation and/or evolution seem to have taken
quiet different routes.

DI\,Her and NY\,Cep are currently the only early-type detached
binaries for which the orientation of the stellar spin axes is known.
Both systems are on eccentric orbits, have detached components, and
are young enough to exclude any development of the spin-orbit
alignment during main-sequence life due to tidal interactions. Yet the
axes in the DI\,Her system are strongly misaligned ($\beta_{\rm
  p}=72\pm4$; $\beta_{\rm s}=-84\pm8$) while the primary axis in
NY\,Cep is well aligned ($\beta_{\rm p}=2\pm4$). Indeed, the projected
alignment of NY\,Cep is even closer than the alignment between the
Sun's spin axis and the ecliptic plane (7$^{\circ}$).

As of yet there is no good explanation for the misalignment of
DI\,Her. Broadly speaking, it could be primordial, or it could be the
result of an interaction between the stars or between the stars and
the disk from which they formed, or it could be due to torques from
additional bodies in the system. The results in this paper do not
point to a particular mechanism, but they do mean that any successful
theory must pass the ``NY\,Cep'' test, namely, it must explain the
difference between those two systems.

Obtaining measurements of spin orbit angles in a number of early type
close binary systems might reveal which environmental variables
determine the orientation of stellar rotation axes in these systems.
For example if the interaction with a distant companion via the Kozai
migration \citep{fabrycky2007} formed close binaries, than misaligned
systems should be more likely to have a detected companion and closer
systems might be more likely to have misaligned axes than wider
systems. If alignment is a simple function of coherence length during
star formation, then the opposite should be true: the closer the
system, the stronger the tendency toward alignment.

This is why we have started a small survey to measure the orientation
of stellar spin axes in other early-type binaries.

\acknowledgments

We are grateful to Greg Henry for his encouragement, and for his
repeated attempts to observe NY\,Cep that were unfortunately foiled by
bad weather. S.A.\ acknowledges support by a Rubicon fellowship from
the Netherlands Organisation for Scientific Research (NWO). J.N.W.\
acknowledges support from a NASA Origins grant (NNX09AD36G). We
acknowledge funding from the Optical Infrared Coordination network
(OPTICON). This research has made use of the Simbad database located
at {\tt http://simbad.u-strasbg.fr/} and the Vienna Atomic Line
database (VALD) located at {\tt http://ams.astro.univie.ac.at/vald/}.

{\it Facilities:} \facility{OHP:1.93m (Sophie)}


\end{document}